\documentclass[aps,prl,reprint,longbibliography,superscriptaddress]{revtex4-2}

\usepackage{graphicx}
\usepackage{dcolumn}
\usepackage{bm}
\usepackage{color}
\usepackage{xcolor}
\usepackage{epstopdf}
\usepackage{gensymb}
\usepackage{ulem}
\usepackage{float}
\usepackage{xr}
\usepackage{sidecap}
\usepackage{mathrsfs}
\usepackage{amsmath}
\usepackage{amsthm}
\usepackage{enumerate}
\usepackage{soul}
\usepackage{amssymb}
\usepackage{units}
\usepackage{comment}
\usepackage{microtype}
\usepackage{cases}
\usepackage{hyperref}

\hypersetup{
    colorlinks=true,
    linkcolor=blue,       
    citecolor=blue,       
    filecolor=blue,       
    urlcolor=blue         
}

\begin{document}
\preprint{APS/123-QED}
\title{Nonlocality-enabled inverse design of Dirac-type and higher-order degeneracies for traveling and evanescent waves in phononic crystals}

\author{Sharat Paul}
\affiliation{Department of Mechanical Engineering, University of Utah, Salt Lake City, UT 84112, USA} 

\author{Md Nahid Hasan}
\affiliation{Department of Mechanical Engineering, University of Utah, Salt Lake City, UT 84112, USA} 
\affiliation{Department of Mechanical Engineering, Montana Technological University, Butte, MT 59701, USA}

\author{Pai Wang}
\thanks{pai.wang@utah.edu}
\affiliation{Department of Mechanical Engineering, University of Utah, Salt Lake City, UT 84112, USA}


\date{\today}
\begin{abstract}
We propose complete tailoring procedures with analytical precision for band degeneracies in one-dimensional (1D) nonlocal phononic crystals, focusing on the role of beyond-nearest-neighbor (BNN) interactions. Unlike trivial Dirac cones at either the center or boundary of Brillouin zone (BZ), we demonstrate non-trivial Dirac-type and higher-order band crossings at any desirable wave number within the BZ by tuning BNN interactions. Our analyses show that odd-indexed BNN interactions determine the quantity and wave number of degeneracy points, while even-indexed BNN interactions primarily affect the frequency. Moreover, we discover new evanescent wave modes and associated degeneracies in the complex-valued wave number. In addition, we study a varieties of spatial-temporal response patterns in time-domain simulations for the interplay between traveling and localized modes at the propagating and evanescent degeneracies. 
\end{abstract}

\maketitle

Dirac cones, characterized by asymptotically linear crossings of dispersion bands, offer unique wave propagation behaviors and potential applications in phononic crystals, vibro-elastic metamaterials, and advanced acoustic devices. They enable exceptional control over traveling waves\,\cite{mei2012first,tian2024hermitian,wan2024deep}, exhibiting phenomena such as defect-insensitive transmission\,\cite{chen2014accidental,li2024utilizing,liang2024acoustic}, constant phase around defects\,\cite{stobbe2019propagation}, energy and acoustic tunneling\,\cite{li2014double,xu2020three,dai2017double,chen2021acoustic,tang2024systematic}, refraction behavior and zero-refractive-index\,\cite{sakoda2012double,yuan2024observation,dai2017double}, robust edge\,\cite{zheng2020acoustic,huang2024realizing,ding2024topological,huang2024topological,gou2024observation,chu2024dual,liang2024acoustic,lu2014dirac,gulzari2024valley,PhysRevLett.115.104302} and corner states\,\cite{zhang2024higher,fan2024multi,gao2024acoustic} with topological phase transitions\,
\cite{chen2016tunable,zheng2020acoustic,chen2022topology,li2024emergence,liu2024higher,okugawa2014dispersion,hu2025unconventional,ryu2024exceptional,2503.03314}, and backward-wave behavior\,\cite{gross2024dispersion}. Additionally, they achieve wave separation\,\cite{wu2024elastic} for energy harvesting\,\cite{dong2024topological} and distribution\,\cite{deng2024asymmetric,liang2024low}, improved waveguiding\,\cite{hu2024robust,xu2024type,10.1063/5.0203024}, ultrasonic nondestructive testing\,\cite{li2024robust}, and topological black hole effect\,\cite{indaleeb2024spin}. 
Dirac-like cones can also lead to exotic wave behaviors, such as pseudo-zero group velocity modes~\cite{lanoy2020dirac}, waves with infinite phase velocity and finite group velocity~\cite{stobbe2017conical,stobbe2017dirac},  Jackiw-Rebbi-type Dirac boundary~\cite{PhysRevLett.129.135501}, and stable degeneracy formations~\cite{kuznetsov2024dirac,lu2014dirac}. Exploiting accidental degeneracy also allows isotropic Dirac cones in multi-dimensional metamaterials\,\cite{sakoda2012dirac}, forming conical points\,\cite{maznev2014dirac}, as well as dipole- and quadruple-degenerate states\,\cite{liu2011dirac,chen2014accidental}.

Historically, Dirac cones were first predicted theoretically in graphene and later experimentally confirmed, showing their unique linear energy-momentum relationship\,\cite{dirac1981principles,PhysRev.71.622,novoselov2005two}. In phononic crystals, Dirac cones were investigated using first-principles methods\,\cite{mei2012first} and confirmed by experiments in honeycomb lattices and triangular arrays\,\cite{liu2011dirac,chen2014accidental}. Fundamental design strategies include symmetry tuning\,\cite{lu2014dirac}, accidental degeneracy, topology optimization\,\cite{DONG2021115687} and adjusting lattice constants to introduce or modify Dirac cones\,\cite{dai2017double,ZHANG2018423}. Incorporating nonlocality\,\cite{chen2021roton,arash2023drawing,DALPOGGETTO2024109503,2410.17329,ONGARO2025110095} further enhances the design and optimization of metamaterials with Dirac cones\,\cite{sinha2024effect,wan2024deep}. While numerous studies have explored Dirac cones in phononic crystals, a fundamental and comprehensive investigation into their formation is still missing, as most research primarily focused on forward-problem methodologies from the lattice structures. 

In this paper, we employ beyond-nearest-neighbor (BNN) interactions~\cite{moore2023acoustic,bossart2023extreme,rajabpoor2023breakdown,chaplain2023reconfigurable,edge2025discrete,PhysRevB.110.144304} to achieve an inverse-design protocol for band degeneracies at any frequency and any wave number, investigating both linear (i.e. Dirac-type) and higher-order (both quadratic and cubic) degeneracies 
of one-dimensional (1D) nonlocal phononic crystals. 
Furthermore, we study a new family of band degeneracies with complex-valued wave number for evanescent waves. In addition, we demonstrate different types of Dirac cones and the time-domain wave dynamics associated with them.

\begin{figure}[b!]
    \centering
    \includegraphics[width=\linewidth]{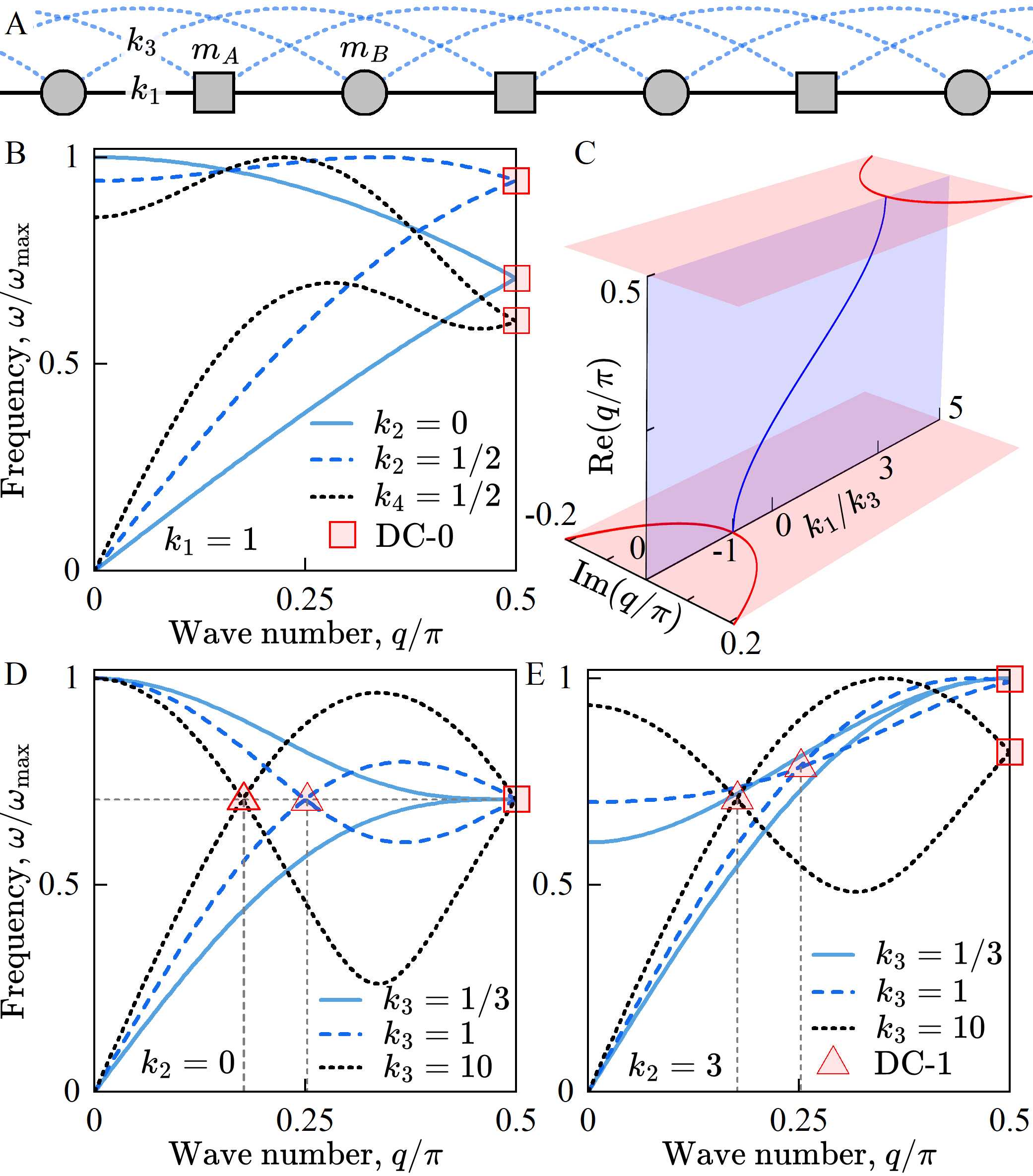}
    \caption{
    (A) Di-atomic nonlocal phononic crystal with masses $m_\textrm{A}$ (square) and $m_\textrm{B}$ (circle), where $k_1$ (black lines) and $k_3$ (blue lines) 
    represent interactions between 1st- and 3rd- nearest neighbors, respectively. 
    (B) Dispersion relations of a lattice with only $k_1$ as the odd-indexed stiffness, which always results in a trivial Dirac cone (DC-0, red square) at the BZ boundary. 
    (C) Real (blue line) and complex (red lines) Dirac cone wave number ($q_\textrm{D}$) versus the stiffness ratio ($k_1/k_3$), as given in Eq.\,\eqref{dirac_soln_N=3}. 
    (D) Dispersion relations for a lattice with $k_1$ and $k_3$ only. 
    The presence of $k_3$ leads to a non-trivial Dirac cone (DC-1, red triangle) 
    within the BZ interior. (E) Demonstrations of the effect of even-indexed $k_n$ on the normalized frequency of the Dirac point. All stiffness values are normalized by $k_1$.}
    \label{fig:DC_soln_N_1_3}
\end{figure}

We start with the general formulation of the one-dimensional diatomic nonlocal phononic lattice, consisting of masses $m_\textrm{A}$ and $m_\textrm{B}$ connected by linear springs $k_n$ where $n = 1, 2, ..., N$. Here, $k_1$ denotes the local stiffness connecting nearest-neighbor masses, while $k_N$ represents the longest-range interaction in the lattice, and $N$ is a proper indicator of the lattice complexity in general. A schematic of the lattice is shown in Fig.\,\ref{fig:DC_soln_N_1_3}A. We note that odd-indexed $k_n$'s connect different types of masses ($m_\textrm{A}$ to $m_\textrm{B}$ and $m_\textrm{B}$ to $m_\textrm{A}$), while even-indexed $k_n$'s connect the same type of masses ($m_\textrm{A}$ to $m_\textrm{A}$ and $m_\textrm{B}$ to $m_\textrm{B}$).  
The dispersion relation is given by 
\begin{align}\label{dispersion_relation_1D}
    \omega_\pm^2(q) = &\Big(\frac{K_0}{m_\textrm{A}}+\frac{K_0}{m_\textrm{B}}\Big) \notag\\ &\pm\sqrt{\Big(\frac{K_0}{m_\textrm{A}}+\frac{K_0}{m_\textrm{B}}\Big)^2+\frac{1}{m_\textrm{A}m_\textrm{B}}\Big(K_1^2 - 4K_0^2\Big)},
\end{align}
where
\begin{equation}
        \label{K0K1_exp}K_0 = \sum_{n=1}^{N}k_n - \sum_{\substack{n=2 \\ \text{even}}}^{N}k_n\cos{(nq)},\quad
    K_1 =2\sum_{\substack{n=1 \\ \text{odd}}}^{N}k_n\cos{(nq)}.
\end{equation}
Here, $\omega$ denotes the frequency, and $q$ is the dimensionless wave number (that is, the product of wave number and lattice constant). 
Eq.\,\eqref{dispersion_relation_1D} provides both the first ($\omega_-$) and second ($\omega_+$) bands of the system. Degeneracy occurs when the two bands cross each other. 
At the crossing point, both bands reach the same frequency at the same wave number $q_\textrm{D}$, 
which provides the following criteria for band crossing at finite frequency:\\ 
\begin{subnumcases}%
    \,m_\textrm{A} = m_\textrm{B}, \label{dirac_condition1} \\
    K_1(q_\textrm{D}) = 0, \label{dirac_condition2} \\ 
    K_0(q_\textrm{D}) > 0. \label{dirac_condition3}
\end{subnumcases}

In Eq.\,\eqref{K0K1_exp}, $K_1(q)$ contains only the odd-indexed stiffness terms $k_1$, $k_3$, $k_5, ...$. Thus, according to Eq.\,\eqref{dirac_condition2}, the odd-indexed $k_n$ terms determine the wave number of the Dirac point, while the even-indexed $k_n$ terms change the frequency. 
The number of Dirac cones depends on the longest-range odd-indexed stiffness, $N_\textrm{odd}=1,3,5, ...$, which results in at most $(N_\textrm{odd}+1)/2$ degeneracies with a trivial Dirac cone always appearing at the Brillouin zone (BZ) boundary, leaving a maximum of $(N_\textrm{odd}-1)/2$ non-trivial crossings at the interior of the positive half of the first BZ. 

For the simplest case without any odd-indexed BNN interactions, we combine $N_\textrm{odd}=1$ with Eq.\,\eqref{dirac_condition2} and obtain the wave number of the Dirac cone as
\begin{equation}\label{dirac_soln_BZ_bound}
    q_\textrm{D} = \frac{\pi}{2}(2z+1), \quad z\in\mathbb{Z}.
\end{equation}
The solutions of Eq.\,\eqref{dirac_soln_BZ_bound} can only admit trivial Dirac cones at the BZ boundary, $q=\pi/2$. 
We illustrate three examples of trivial Dirac cones in Fig.\,\ref{fig:DC_soln_N_1_3}B, where all lattices have only one odd-indexed stiffness, $k_1$ (i.e. the local connection) but may include even-indexed BNNs, such as $k_2$ and $k_4$. Here, we 
normalize frequencies in the plot with the maximum cutoff frequency of the each case.  
While the examples with $k_2=(1/2)k_1$, and $k_4=(1/2)k_1$ do exhibit non-monotonic dispersion bands, only a trivial Dirac cone is possible here.

Next, we focus on the emergence of non-trivial Dirac cones. 
With lattice complexity $N=3$, Eq.\,\eqref{dirac_condition2} gives
\begin{equation}\label{dirac_soln_N=3}
    q_\textrm{D} = \pm\cos^{-1}{\Big(\pm\sqrt{\frac{3}{4} - \frac{k_1}{4k_3}}\Big)}+2z\pi, z\in\mathbb{Z}, 
\end{equation}
yielding a maximum of two Dirac cones in the positive half of the first BZ: a trivial one (DC-0) and a non-trivial one (DC-1). 
To achieve a non-trivial Dirac cone at a real-valued wave number, Eq.\,\eqref{dirac_soln_N=3} gives that $k_1/k_3$ needs to be within the range of $[-1, 3]$. We illustrate this fact in Fig.\,\ref{fig:DC_soln_N_1_3}C, where the blue and red lines represent the real- and complex-valued wave number of the non-trivial Dirac cone, $q_\textrm{D}$, respectively. 

In the meantime, the inequality of (\ref{dirac_condition3}) provides an additional constraint,\\ 
\begin{equation}\label{even_k_for_N=3}
    k_2 > \frac{k_1 + k_3}{\cos{(2q_\textrm{D})} - 1},
\end{equation}
which offers the range of valid $k_2/k_1$ for any specific $k_3/k_1$. We can achieve valid dispersion relations with desirable crossing by also requiring Im$\{\omega\}=0$ at all wave numbers. Fig.\,\ref{fig:DC_soln_N_1_3}D shows that the Dirac point changes its wave number, $q_\textrm{D}$ with different $k_3$. In most cases, as exemplified by $k_3=k_1$ and $10k_1$, we get one trivial and one non-trivial Dirac cones. However, at $k_3=(1/3)k_1$, the two Dirac cones merge together giving rise to a higher-order 
crossing\,\cite{PhysRevB.110.L121122} illustrated by the solid blue curves in Figs.\,\ref{fig:DC_soln_N_1_3}D and \,\ref{fig:DC_soln_N_1_3}E.

We now outline a protocol to realize a non-trivial Dirac point at any frequency $\omega_\textrm{Target}$, and any wave number $q_\textrm{Target}$, using the nonlocal phononic crystal with $N=3$: First, we use $q_\textrm{D} = q_\textrm{Target}$ with Eq.\,\eqref{dirac_soln_N=3} to solve for the ratio of $k_3/k_1$. Then, using $\omega_+ = \omega_- = \omega_\textrm{Target}$ with Eq.\,\eqref{dispersion_relation_1D} we can obtain the ratio of $k_2/k_1$. This two-step procedure can guarantee the emergence of an arbitrarily specified band crossing.

To realize multiple non-trivial Dirac points, we need more complex lattice designs with $N_\textrm{odd}>3$, 
where $N_\textrm{odd}$ is the longest-range odd-indexed interactions in the lattice. To demonstrate the existence of more than one non-trivial Dirac cone, we analyze the case of $N=5$. This allows for a maximum of three Dirac cones, one trivial (DC-0) 
and two non-trivial (DC-1 and DC-2), to exist simultaneously. The non-trivial ones are at 
\begin{equation}\label{dirac_soln_N=5}
    q_\textrm{D} = \pm\cos^{-1}\left(\pm\sqrt{\kappa}\right) + 2z\pi, \quad z \in \mathbb{Z},
\end{equation}
where
\begin{equation}\label{y_for_N=5}
    \kappa = -\Big(\frac{k_3}{8k_5} - \frac{5}{8}\Big)\pm\sqrt{\frac{1}{64}\Big(\frac{k_3^2}{k_5^2}+2\frac{k_3}{k_5}-4\frac{k_1}{k_5}+5\Big)},
\end{equation}
and $0\leq \kappa \leq 1$ for $q_\textrm{D}$ to be real-valued. It then follows that we need 
\begin{equation}\label{k3_k5_N5}
    \frac{k_3^2}{k_5^2}+2\frac{k_3}{k_5}-4\frac{k_1}{k_5}+5\geq 0,
\end{equation}
which provides the admissible parameter space of the odd-indexed $k_n$'s. Moreover, using the inequality (\ref{dirac_condition3}), we find the constraint for even-indexed $k_n$'s as,
\begin{align}
    \frac{k_2}{k_1} + \frac{k_4}{k_1} - \frac{k_2}{k_1}\cos{(2q_\textrm{D})} - \frac{k_4}{k_1}\cos{(4q_\textrm{D})}> -1 - \frac{k_3}{k_1} - \frac{k_5}{k_1},
\end{align}

\begin{figure}[b!]
    \centering
    \includegraphics[width=\linewidth]{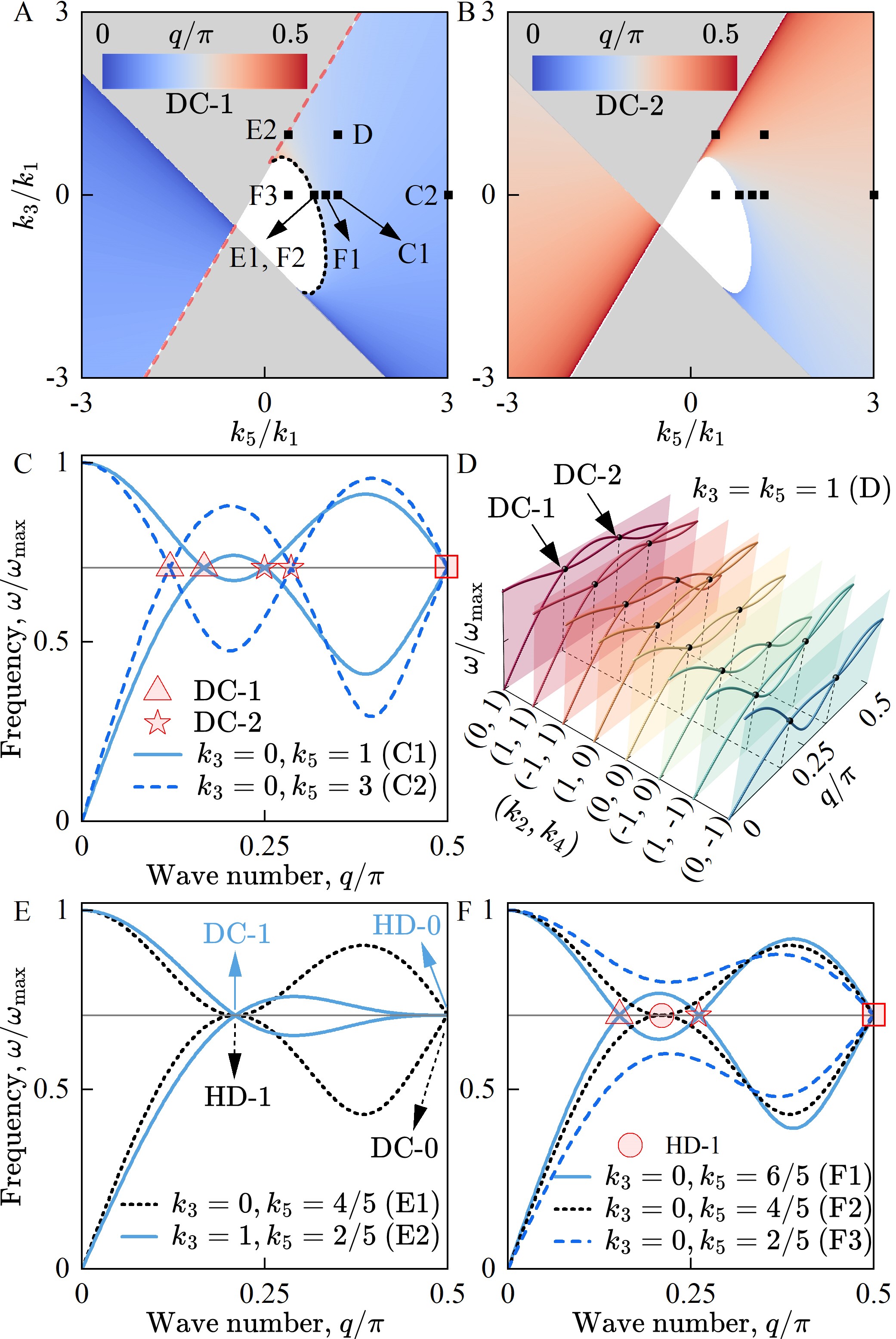}
    \caption{
    Band degeneracies for the lattice with $N=5$: 
    (A) and (B) show how the wave numbers of non-trivial Dirac cones (DC-1 and DC-2) depend on the stiffness ratios. While the trivial Dirac point (DC-0) always exists, the color-gradient regions admit two additional non-trivial Dirac cones, and the gray areas has one non-trivial Dirac cone. The black square dots correspond to stiffness ratios used in the dispersion bands shown in (C)–(F). (C) Dispersion relations of two lattices with different $k_5$, showing that the odd-indexed stiffness affects the wave number of the non-trivial Dirac cones while keeping the normalized frequency constant. Here, the triangle and star indicate the first (DC-1) and second (DC-2) non-trivial Dirac cones, respectively. (D) Selected cases illustrating the effect of even-indexed stiffness ($k_2$ and $k_4$) while keeping $k_3$ and $k_5$ constant. (E) Cases with a non-trivial higher-order degeneracy (HD-1 on E1) as shown on the dotted black line dispersions and with a trivial higher-order degeneracy (HD-0 on E2) at the BZ boundary shown on the solid blue line dispersions. (F) Dispersion relations showing the merging of two non-trivial Dirac cones into a higher-order crossing and their subsequent separation through tuning of $k_5$. Here, the red circle indicates the non-trivial higher-order degeneracy (HD-1).}
    \label{fig:q_soln_N_5}
\end{figure}

We show the parameter space of $k_3/k_1$ and $k_5/k_1$ for two non-trivial Dirac cones in Figs.\,\ref{fig:q_soln_N_5}A and \ref{fig:q_soln_N_5}B in the range of -3 to 3. 
Each point in the colored region represents a case with two non-trivial Dirac cones. As the color maps show, depending on the ratio of $k_3/k_5$, DC-1 and DC-2 can appear with any wave number $0<q_\textrm{D}<\pi/2$. 
In contrast, each point in the gray area offers only one non-trivial crossing together with the trivial DC-0, while the white region at the center means DC-0 only.
Moreover, at the boundaries of the colored regions, we have the cases of two Dirac cones merging into one higher-order degeneracy point.  

Within the parameter range of $k_3$ and $k_5$ in Figs.\,\ref{fig:q_soln_N_5}A \& \ref{fig:q_soln_N_5}B, we select a few examples to demonstrate the properties of non-trivial degeneracies. The points C1 and C2 in Fig.\,\ref{fig:q_soln_N_5}A share the same $k_3 = 0$ but have different $k_5=k_1$ and $k_5=3k_1$, respectively. The corresponding dispersion curves in Fig.\,\ref{fig:q_soln_N_5}C show that the two non-trivial DC-1 (triangle mark) and DC-2 (star mark) are moving apart from each other with increasing $k_5$, while the trivial DC-0 (rectangle mark) remains at the BZ boundary. 
Next, at point D in Fig.\,\ref{fig:q_soln_N_5}A, we fix $k_5 = k_3=k_1$, and then vary the even-indexed $k_2$ and $k_4$. Several corresponding dispersion curves are depicted in Fig.\,\ref{fig:q_soln_N_5}D. 
As dictated by Eq.\,\eqref{dirac_soln_N=5}, DC-1 and DC-2 in all these cases are always at $q=\pi/6$ and $q=\pi/3$, respectively, and the even-indexed stiffness values influence the frequencies of the degeneracy points but not the wave numbers. 
Also marked in Fig.\,\ref{fig:q_soln_N_5}A, at the points on the curved boundary (dotted black line), such as E1, two non-trivial Dirac points merge together to create a non-trivial higher-order (i.e. locally asymptotically nonlinear) degeneracy HD-1 at the BZ interior. In contrast, at the points on the straight boundary (red dashed line) in Fig.\,\ref{fig:q_soln_N_5}A, such as E2, a non-trivial Dirac point merges with the trivial Dirac point to create a trivial higher-order degeneracy HD-0 at the BZ boundary. The corresponding dispersion curves showing those quadratic crossings are depicted in Fig.\,\ref{fig:q_soln_N_5}E. 
Our analysis also show that, while the E1 dispersion curves are locally quadratic at HD-1, the E2 dispersion curves are locally cubic at HD-0. This enriches the varieties of degeneracies that are possible in nonlocal phononic crystals. 
Lastly, we focus on tuning only $k_5/k_1$ while setting all other $k_n$'s to zero. In Fig.\,\ref{fig:q_soln_N_5}F, three dispersion curves corresponding to Points F1, F2 and F3 demonstrate the transition from two non-trivial Dirac cones (F1) to one non-trivial quadratic crossing (F2), and then to the trivial Dirac cone only (F3). 
\begin{figure*}
    \centering
    \includegraphics[width=\linewidth]{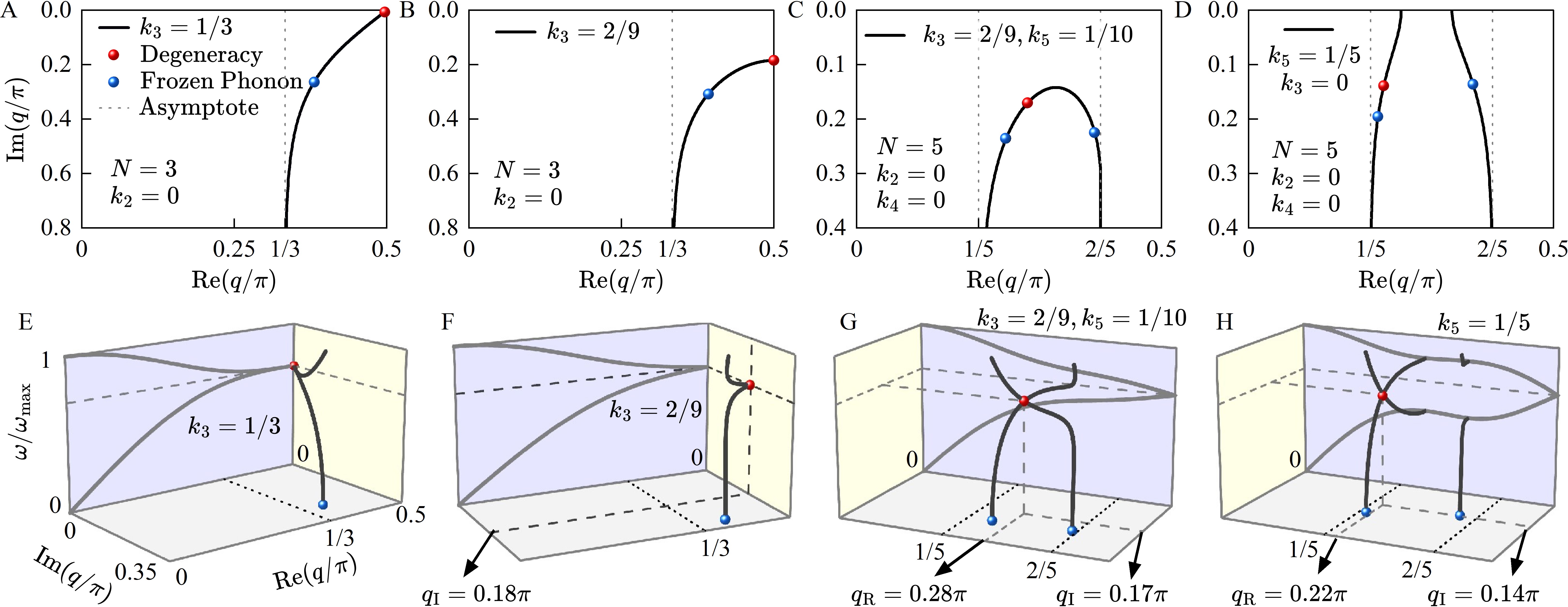}
    \caption{
    Evanescent band degeneracies with complex-valued wave numbers. (A)-(D) show the wave number solutions of Im$\{\omega\}=0$ for evanescent modes. These black lines represent the projection of the evanescent modes onto the complex wave-number plane. In all cases, the solutions exhibit asymptotic behavior, indicated by dotted vertical lines. 
    The red and blue dots indicate locations of band degeneracies and the frozen phonons, respectively. (A) With $k_3 = (1/3)k_1$, the degeneracy occurs at $q=\pi/2$, and the frozen phonon occurs at $q=(0.38\pm i0.26)\pi$. 
    (B) With $k_3=(2/9)k_1$, the evanescent degeneracy is at $q=(0.5\pm i0.18)\pi$. (C) The addition of $k_5=(1/10)k_1$ creates two asymptotes at $q_\textrm{R}=\pi/5$ and $2\pi/5$. The non-trivial evanescent degeneracy is at $q=(0.28\pm i0.17)\pi$. 
    (D) A lattice with $k_5=(1/5)k_1$ and $k_3=0$ also provides two asymptotes and put the evanescent degeneracy at a smaller imaginary wave number, $q=(0.22\pm i0.14)\pi$. (E)-(H) shows the corresponding evanescent band structures. The gray curves represent the propagating bands, and the black curves represent the evanescent bands. 
    (E) For $k_3 = (1/3)k_1$, all propagating and evanescent bands meet at a single degeneracy point. (F) The stronger $k_3=(2/9)k_1$ separates the degeneracy points for propagating and evanescent bands. 
    (G) The addition of $k_5$ affects the wave number of the evanescent degeneracy. (H) With $k_5=(1/5)k_1$ as the only nonlocal stiffness, the non-trivial evanescent degeneracy also exists.}
    \label{fig:Imaginary_crossing}
\end{figure*}

Next, we extend our study to band degeneracies in evanescent waves\,\cite{chen2024anomalous} 
by introducing the complex-valued wave number $q=q_\textrm{R}+iq_\textrm{I}$. 
We focus on lattices with odd-indexed $k_n$'s and require that the frequency is real valued. The condition Im\{$\omega(q)\}=0$ entails 
\begin{equation}\label{q_C_for_evan_mode}
    \sum_{\substack{n=1 \\ \text{odd}}}^{N} k_n\sin{(nq_\textrm{R})} \sinh{(nq_\textrm{I})} = 0.
\end{equation}
For local lattices ($N=1$), Eq.\,\eqref{q_C_for_evan_mode} is simply 
\begin{equation}
    k_1\sin{(q_\textrm{R})}\sinh{(q_\textrm{I})} = 0, 
\end{equation}
which means that evanescent modes are possible at $q_\textrm{R}=0$ only for conventional phononic crystals with local nearest-neighbor interactions only. By adding $k_3$ (i.e., using nonlocal lattices with $N=3$), Eq.\,\eqref{q_C_for_evan_mode} becomes 
\begin{equation}\label{Im_crossiong_N=3}
q_\textrm{I}=\sinh^{-1}\left({\sqrt{\frac{12k_3\sin^2(q_\textrm{R}) - (k_1+9k_3)}{12k_3 - 16k_3\sin^2{(q_\textrm{R})}}}}\right),
\end{equation}
which provides the solution pairs of $q_\textrm{R}$ and $q_\textrm{I}$, for which the evanescent mode exists. We note that these solutions form an asymptote at $q_\textrm{R}=\pi/3$ for $N=3$. More generally, through singularity analysis for any $N$, we find that the number of asymptotes depends on the longest-range odd index stiffness $N_\text{odd}$: A total of $[N_\text{odd}-1]/2$ asymptotes exist, and they are at $q_\textrm{R} =n\pi/N$ for $n\in \{1,2,3,\ldots,N\}$. For example, $N=5$ entails two asymptotes at $q_\textrm{R}=\pi/5$ and $2\pi/5$, while $N=7$ leads to three asymptotes at $q_\textrm{R} = \pi/7$, $2\pi/7$, and $3\pi/7$. 

To illustrate these asymptotic behaviors, in Figs.\,\ref{fig:Imaginary_crossing}A and \ref{fig:Imaginary_crossing}B, we plot the solutions of Eq.\,\eqref{Im_crossiong_N=3} for $k_3=(1/3)k_1$ and $k_3=(2/9)k_1$, respectively. We note that the solution curves here are simply the projections of the evanescent wave dispersion curves (both bands) on the complex plane of wave numbers. A lattice with complexity $N=3$ gives rise to one asymptote at $q_\textrm{R}=\pi/3$ shown as dotted vertical lines in Figs.\,\ref{fig:Imaginary_crossing}A and \ref{fig:Imaginary_crossing}B. 
Furthermore, we also numerically solve Eq.\,\eqref{q_C_for_evan_mode} with $N=5$. Figure.\,\ref{fig:Imaginary_crossing}C shows the effect of introducing a $k_5= (1/10)k_1$ together with $k_3=(2/9)k_1$, 
while Fig.\,\ref{fig:Imaginary_crossing}D illustrates the solution in the case of $k_5=(1/5)k_1$ and $k_3=0$. As predicted, both cases of $N=5$ show two asymptotes at $q_\textrm{R}=\pi/5$ and $q_\textrm{R}=2\pi/5$, respectively. 

Corresponding to the solution distributions in Figs.\,\ref{fig:Imaginary_crossing}A-\ref{fig:Imaginary_crossing}D, the black curves in the three-dimensional ($q_\textrm{R}, q_\textrm{I}, \omega$) space in Figs.\,\ref{fig:Imaginary_crossing}E-\ref{fig:Imaginary_crossing}H represent the bands of evanescent waves. Here, the gray curves on the ($q_\textrm{R}, \omega$)-plane, where $q_\textrm{I}=0$, are the bands of traveling waves in the same lattice. We use the red circular dots to highlight the focus of our study — the degenerate points of evanescent waves, while the blue circular dots represent ``frozen phonons"~\cite{PhysRevLett.134.086101,chen2024anomalous} at zero frequency. 
In particular, Figure.\,\ref{fig:Imaginary_crossing}E illustrates dispersion relations with complex-valued wave numbers for the lattice with $k_3=(1/3)k_1$. It shows that the traveling and evanescent bands all merge together at a single crossing point at $(q_\textrm{R},q_\textrm{I})=(\pi/2,0)$, where the group velocity $v = \partial\omega / \partial q_\textrm{R}$ vanishes for all modes. Although we do not have a well-defined decay length (since $q_\textrm{I}=0$) as that for other evanescent wave modes, this mode still can not propagate in the chain due to the zero group velocity (ZGV). Furthermore, we also note that this mode is more strongly localized than typical ZGV modes like rotons~\cite{chen2021roton} 
or maxons~\cite{10.1063/5.0179959,10.1063/5.0180074}
, because the second derivative $\partial^2\omega/\partial q_\textrm{R}^2$ also vanishes for all bands at this point. 
It corresponds to the undulation point~\cite{arash2023drawing}
, stationary inflection point, or higher-order van Hoove singularity~\cite{PhysRevB.101.125120} that are of interest in a varieties of systems. 
In the context of phononic crystals, it manifests itself as a excitable mode that not only non-propagating but also non-spreading.  

In contrast, Figure.\,\ref{fig:Imaginary_crossing}F shows the case with $k_3=(2/9)k_1$. The ``evanescent degeneracy" is now at a complex-valued wave number, $q=(0.5\pm i0.18)\pi$, that is separate from the trivial Dirac cone at $q=\pi/2$ for traveling waves. To the best of our knowledge, this is the first time that an evanescent band crossing is studied in phononic crystals, and this new phenomenon might open new avenues for future research. 
Our analysis also shows that lattices with complexity $N=3$ can give rise  one evanescent degeneracy at $q_\textrm{R}=\pi/2$ only. 
To overcome this limitation and expand of design freedom, we add $k_5$ and show the results in Figs.\,\ref{fig:Imaginary_crossing}G and \ref{fig:Imaginary_crossing}H, 
where the evanescent degeneracy appears at $q=(0.28\pm i0.17)\pi$ and $q=(0.22\pm i0.14)\pi$, respectively. In each case, we also have two frozen phonons at zero frequency.

To demonstrate wave behaviors of different degenerate modes, we perform time-domain simulations on finite lattices with the excitation force of the form,  
\begin{equation}
    f(t,x)=f_0 \exp[{-(t_\textrm{m} - t)^2/\tau^2}] \cos{(\omega_\textrm{c} t)},
\end{equation}
where $f_0$ is the amplitude, $\omega_\textrm{c}$ is the prescribed carrier frequency, $t_\textrm{m}$ and $\tau=50/\omega_\textrm{c}$ are the mean and span of the gaussian envelope. 
\begin{figure}
    \centering
    \includegraphics[width=\linewidth]{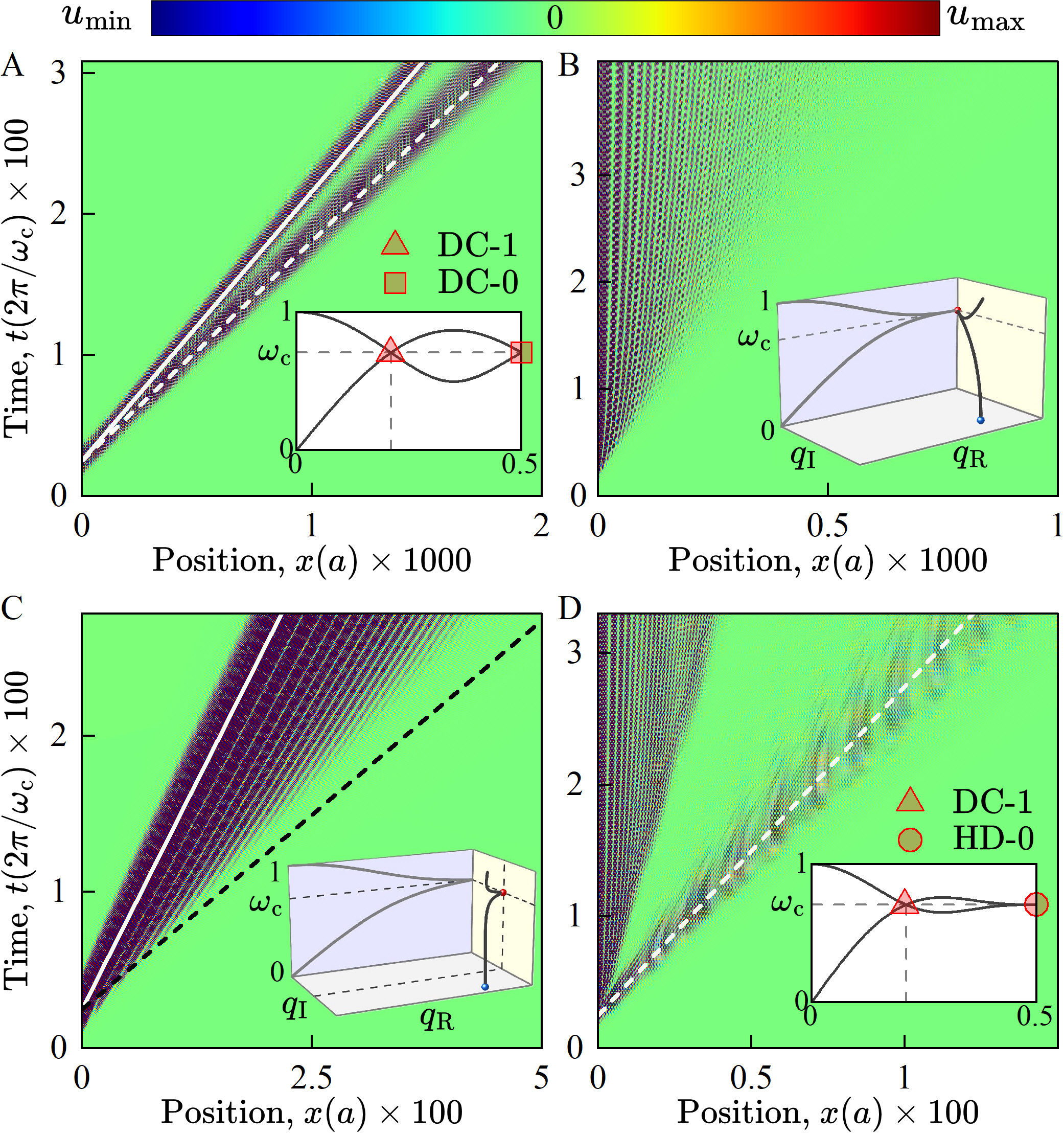}
    \caption{
    Time-domain simulation results of four example lattices with corresponding dispersion bands shown as insets: 
    (A) For a lattice with $k_2=0$ and $k_3=2k_1$, a non-trivial Dirac cone at $q_\textrm{DC-1}=0.21\pi$ and a trivial Dirac cone at $q_\textrm{DC-0}=\pi/2$ have same frequency. The time-domain response to a Gaussian wave packet with this frequency is plotted. The solid and dotted white lines represent the group velocities at DC-0 and DC-1, respectively. (B) Time-domain response of the higher-order degeneracy at $q_\textrm{R}=\pi/2$ (shown in Fig.\,\ref{fig:Imaginary_crossing}E) is plotted in the 
    lattice with $k_3=(1/3)k_1$. (C) For the lattice with $k_3=(2/9)k_1$, two Dirac cones, one for propagating mode, the other for evanescent mode, share the same frequency (shown in Fig.\,\ref{fig:Imaginary_crossing}F). The white solid line corresponds to the group velocity of the propagating Dirac cone, while the dashed black line represents the group velocity 
    of the evanescent degeneracy. (D) For the lattice with $k_3=k_1$ and $k_5=(2/5)k_1$, the Dirac cone at $q=0.21\pi$ and the higher-order crossing (ZGV) at $q=\pi/2$ (E2 curves in Fig.\,\ref{fig:q_soln_N_5}E) share the same frequency. The white dashed line corresponds to the group velocity of at the Dirac cone. 
    } 
    \label{fig:time-domain}
\end{figure}
As the first example, Figure.\,\ref{fig:time-domain}A illustrates the time-domain simulation results of a lattice with $k_3=2k_1$ and $k_2=0$. We excite one mass in the middle of the 1D chain containing 4000 masses and plot the space-time response of the right half of the chain, as the results are symmetric on both sides. The inset of the figure shows the corresponding band structure, which exhibits a non-trivial Dirac cone at $q_\textrm{DC-1}=0.21\pi$ and a trivial Dirac cone at $q_\textrm{DC-0}=\pi/2$. Both Dirac cones are at the same frequency, $\omega=0.71\omega_\text{max}$, and each is a type-I Dirac cone (discussed in the following section) with symmetric group velocities: $|v_\textrm{DC-1}|=0.79$ and $|v_\textrm{DC-0}|=0.65$, which are depicted as the slopes of the dashed and solid white lines, respectively. Their values are calculated directly from the dispersion relations as 
\begin{equation}\label{group_velocity}
    v_\pm = \frac{\partial \omega_\pm}{\partial q_\textrm{R}} = \frac{1}{2m\omega_\pm}\left(2\frac{\partial K_0}{\partial q_\textrm{R}}\pm\frac{\partial K_1}{\partial q_\textrm{R}}\right).
\end{equation}
Figure \,\ref{fig:time-domain}B shows the results for an excitation of at the band-crossing frequency of Fig.\,\ref{fig:Imaginary_crossing}E, where the quadratic band crossing coexists both in real and evanescent wave modes. All modes at this frequency, $\omega=0.71\omega_\text{max}$, exhibit zero group velocity (ZGV). Though we implement a chain with 5000 masses for the simulation, we show results for only 1000 masses because the only mode here is non-propagating. 
The spreading behavior apparent in the space-time plot is due to higher-order diffusion, not wave propagation. 
In Fig.\,\ref{fig:time-domain}C, we plot results of an excitation of at the Dirac point frequency of Fig.\,\ref{fig:Imaginary_crossing}E. Here we have two separate crossing points, one for traveling waves with a real wave number $q=\pi/2$ (DC-R), the other for evanescent waves with a complex wave number $q=(0.5+i0.18)\pi$ (DC-C). They occur at the same frequency $\omega=0.71\omega_\textrm{max}$. 
The slopes of the solid white line and dashed black line indicate the their group velocities, $|v_\textrm{DC-R}|=0.09$ and $|v_\textrm{DC-C}|=0.22$, respectively. As the traveling wave propagates forward, it repeatedly excites the evanescent mode, which has a faster group velocity but also decays exponentially in space due to the non-zero $q_\textrm{I}$. This leads to the fractal-like pattern in the space-time response. We note that fractal and/or self-similar patterns are well known to occur in phase space, and they are also shown to exist in the parameter space \,\cite{PhysRevE.57.1544,PhysRevE.108.L022201,HASAN2025102299} of vibration phenomena. In contrast, the dynamic interplay between propagating and evanescent waves here gives rise to fractal-like patterns in a spatial-temporal plot, and it might leads to new phenomena or functionalities upon further research. 
In Fig.\,\ref{fig:time-domain}D, we show the results of an excitation at the band crossing frequency shown in Fig.\,\ref{fig:q_soln_N_5}E (Specifically, the E2 curve with $k_3=k_1$ and $k_4=0.4 k_1$). Here, we have a Dirac cone at $q_\textrm{DC-1}=0.21\pi$ with the group velocity, $|v_\textrm{DC-1}|=0.44$ and quadratic crossing  at $q_\textrm{HD-0}=\pi/2$ with $|v_\textrm{HD-0}|=0$ (i.e. a ZGV mode). 
As both band crossings occur at the same frequency, the time-domain data shows that the excitation energy splits into two parts. While the traveling wave propagates away from the forcing site, the ZGV mode localizes near the source.

\begin{figure}
    \centering
    \includegraphics[width=\linewidth]{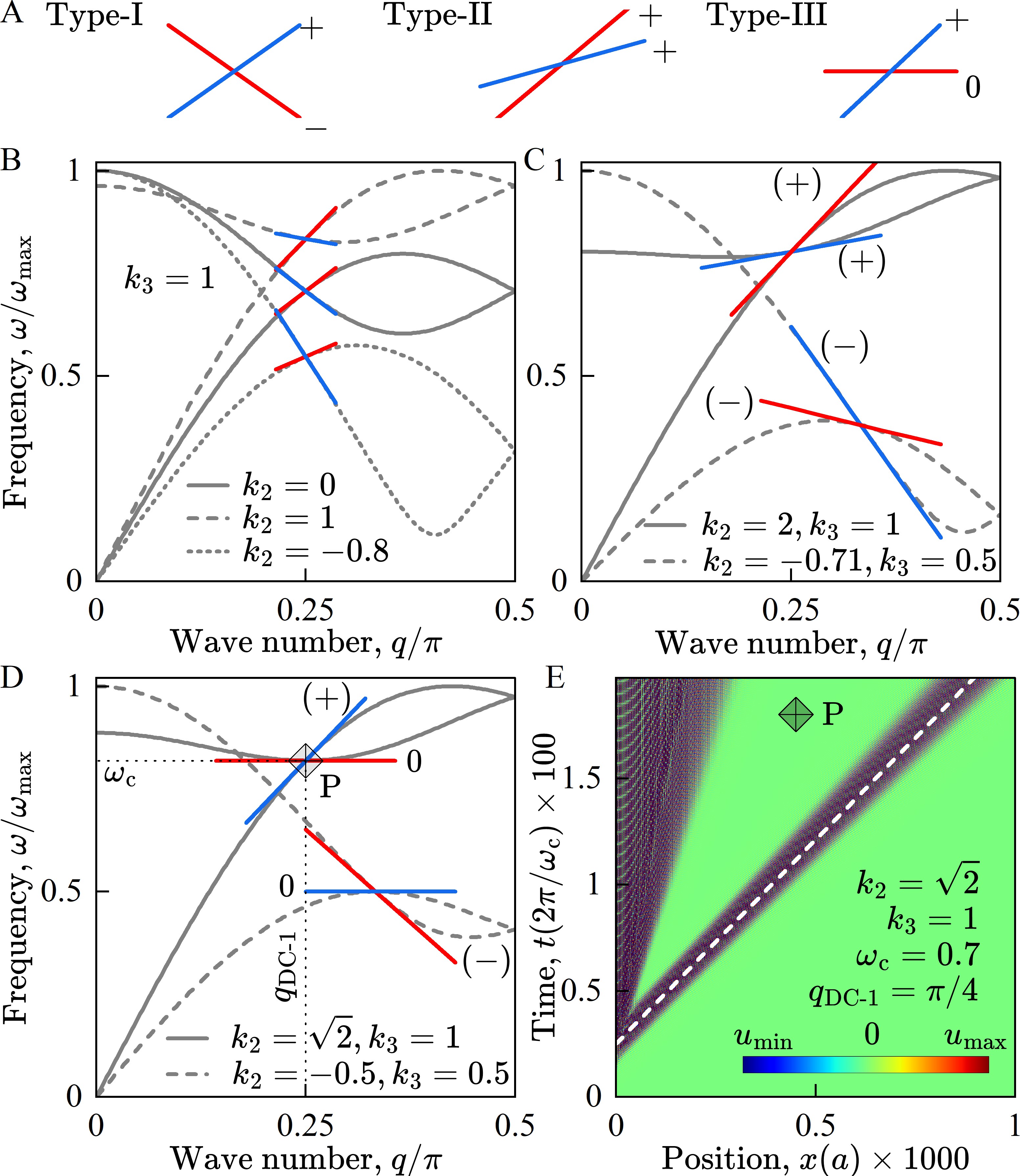}
    \caption{
    Dirac cones of three different types: (A) Dirac points categorized by the signs ($+, -$, or 0) of the group velocities. 
    (B) Type-I Dirac cones, which include symmetric ($k_2=0$) and asymmetric (tilted, $|k_2|>0$) cases. (C) Two kinds of type-II Dirac cones. (D) Type-III Dirac cones with zero group velocity in one band. (E) Time domain analysis of type-III Dirac cones (marked as point P in sub-figure D), showing the coexistence of propagating and non-propagating ZGV modes at the same frequency and wave number
    . The dotted white line indicates the group velocity of the propagating mode.
    }
    \label{fig:dirac_types_N3}
\end{figure}
 Lastly, we use our design protocol to realize different Dirac degeneracies: type I, II, and III, based on the sign of the group velocity of the degenerate modes~\cite{kim2020universal,milicevic2019type,soluyanov2015type,PhysRevLett.117.224301}. The categories are illustrated in Fig.\,\ref{fig:dirac_types_N3}A, in which the line segments represent the local tangents of the dispersion bands at the crossing point. 
To achieve the type-specific customization, we start with Eq.\,\eqref{group_velocity} and focus on the case of $N=3$ to obtain

\begin{align}
    \label{del_w+N3}
    v_\pm &= \frac{1}{\omega_\pm}\Big[\mp k_1\sin{(q)} + 2k_2\sin{(2q)} \mp 3k_3\sin{(3q)}\Big].
\end{align}
As an example, we first choose a desirable wave number for the Dirac point,  $q_\textrm{D}=\pi/4$, which leads to $k_3 = k_1$ for the lattice design and entails the following criteria:
\begin{equation}\label{Eq-type-conditions}
\begin{aligned}
    &\textrm{Type I: } &&|k_2|<\sqrt{2}k_1,\\
    &\textrm{Type II: } &&|k_2|>\sqrt{2}k_1, \\ 
    &\textrm{Type III: } &&|k_2|=\sqrt{2}k_1. 
\end{aligned}
\end{equation}
As another example, if we choose $q_\textrm{D} = \pi/3$, which leads to $k_3 = k_1/2$, then the criteria become $|k_2|<k_1/2, |k_2|>k_1/2$, and $|k_2|=k_1/2$ for Type I, II, and III, respectively. 

In Fig.\,\ref{fig:dirac_types_N3}B, we show the three Type I Dirac crossings at the same wave number, $q_\textrm{D}=\pi/4$. Here, $k_2=0$ provides a symmetric (i.e., same magnitude of the positive and negative group velocities) Type I Dirac point, while $k_2=k_1$ and $k_2=-0.8k_1$ give asymmetric Type I  Dirac points. Fig.\,\ref{fig:dirac_types_N3}C shows two type-II Dirac cones, one at $q_\textrm{D}=\pi/4$, where the group velocities of the two bands at the Dirac point are both positive, and the other at $q_\textrm{D}=\pi/3$, where the group velocities are both negative. In Fig.\,\ref{fig:dirac_types_N3}D, we illustrate two Dirac points of Type III, one with group velocities ($+, 0$) at $q_\textrm{D}=\pi/4$, and the other with group velocities ($-, 0$) at $q_\textrm{D}=\pi/3$, respectively. As type-III Dirac cone provides a propagating mode and a non-propagating (ZGV) mode at not only the same frequency but also the same wave number, we show the corresponding time-domain simulation results in Fig.\,\ref{fig:dirac_types_N3}E for an excitation at point P in Fig.\,\ref{fig:dirac_types_N3}D. 
The white dashed line here indicates the group velocity, $v=0.68$, of the propagating mode, while the ZGV mode is trapped near the excitation site.

In summary, we present a thorough analysis of band crossings enabled by beyond-nearest-neighbor connections in 1D nonlocal phononic crystals, solving the inverse problem of design. Our findings reveal that an arbitrary number of Dirac points are achievable at any desired wave numbers and frequencies, which governed by the odd- and even-indexed stiffness. We explore the merging of Dirac cones, creating quadratic crossing points with non-propagating degenerate modes. Moreover, we also investigate Dirac point of evanescent waves that have complex-valued wave numbers. We provide the analytical closed-form solutions and study wave packet dynamics of those degenerate modes. By tuning stiffness values, we demonstrate control over the types of Dirac cones and uncover fractal-like wave dynamics emerging from the shared frequency of traveling and evanescent Dirac points. These findings hold significant implications for the field of phononic crystals and metamaterials - The ability to precisely design band crossing with detailed customization opens avenues for defect-insensitive wave propagation, robust waveguides, and energy tunneling in advanced acoustic devices. The fractal-like behavior and new evanescent modes could inspire novel applications in wave filtering, sensing, and nondestructive testing. The analytical framework established here also lays the foundation for future exploration of wave dynamics in higher-dimensional systems. This work contributes to the growing understanding of wave manipulation in nonlocal lattices and broadens the potential for engineering innovative phononic and photonic devices.


\section{Acknowledgment} The authors are supported by the startup funds provided by the Department of Mechanical Engineering at University of Utah. This research is also partially supported by the National Science Foundation under Grant No.\,2341003. The support and resources from the Center for High-Performance Computing at the University of Utah are gratefully acknowledged.

\nocite{}
\normalem
\bibliography{apssamp}

\end{document}